\documentclass{iau}
\usepackage{graphicx}

\title[JD 11.~~Panoramic Spectroscopy of Mrk171] %% give here short title %%
{Results on Panoramic Spectroscopy of Mrk171}

\author[S.Hakopian et.al.]   %% give here short author list %%
{ S.A.Hakopian, S.K.Balayan and T.A.Movsessian
 \ } 
\affiliation{Byurakan Astrophysical Observatory,(BAO), Byurakan, 0213, Armenia.
\\ email: {\tt susanaha@bao.sci.am} \\[\affilskip]
{}}
\pubyear{2008}
\volume{xxx}  %% insert here IAU Symposium No.
\pagerange{}
\jname{Title of your IAU Symposium}
\editors{A.C. Editor, B.D. Editor \& C.D. Editor, eds.}
\usepackage{amsmath}
\usepackage{amssymb}

\begin{document}

\maketitle
\begin{abstract}
Observations of Mrk 171, aimed at conduction of panoramic spectroscopy, were undertaken with the Byurakan 2.6-m telescope using spectrograph "VAGR". Within the two components of the galaxy, Mrk171W and Mrk171E, there were differentiated eight condensations of starforming activiity, i.e. HII-regions, and no sign of AGN activity was revealed inspite of existing suggestions.
%% add here a maximum of 10 keywords, to be taken form the file <Keywords.txt>
\end{abstract}

\firstsection % if your document starts with a section,
 % remove some space above using this command.
\section{Introduction}

Mrk 171 is one of the unique nearest galaxies with complicated morphology and richness in physical characteristics. The main part of the galaxy is "pressed" by two big zones of radiation in the radio (\cite [Hibbard \& Yun 1999]{Hibbard Yun99}). The double structure of the galaxy is interpreted as an initial stage of merging processes (e.g.,\cite[Nordgren et al. 1997]{Nordgren et al97}). Among versatile studies are those connected to high quantity of supernovas (e.g.,\cite[Van Dyk et al. 1999]{Van Dyk et al99}), and to the presence of X-ray sources (e.g.,\cite[Heckman et al. 1999]{Heckman_etal99}). Due to its composition of gases, including the molecular CO (e.g., \cite [Sargent \& Scoville 1991]{SargentScovillle91}), it has been studied as a candidate for the galaxies of WR type (e.g., \cite [Ho \& Fillipenko 1995]{HoFillipenko95}). Studies in the ultraviolet and optical ranges are revealing only starformation activity (e.g., \cite [Grimes et al. 2009]{Grimes et al09}), but signs of AGN presence are nevertheless being searched.

\begin{figure}[b]
% \vspace*{}\emph{b}
\begin{center}
 \includegraphics[width=4.8in]{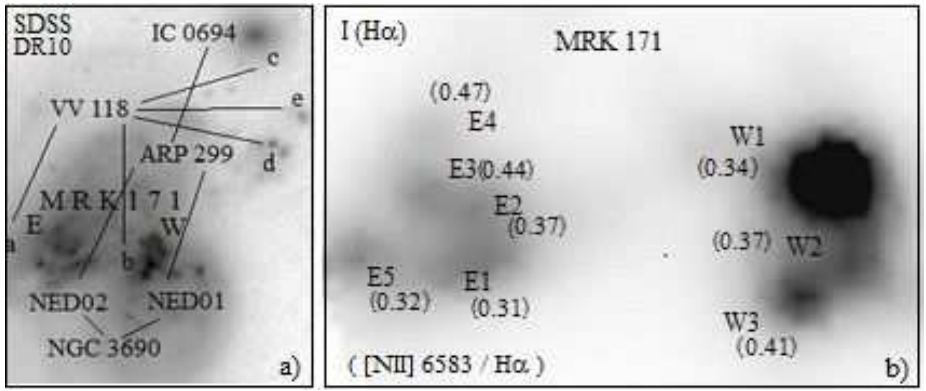} 
% \vspace*{-1.0 cm}
\caption{a) Designations of the components of Mrk171; b )part of Ha-intensity map in smoothed pixels, obtained at 2.6-m telescope of Byurakan with spectrograph "VAGR".}
   \label{fig1}
\end{center}
\end{figure}

\section{Description and Observations}

For the two components of the galaxy we use designations  Mrk171E and Mrk171W, according to their mutual disposition. Along with these, their other designations as components of double galaxy NGC 3690, triple system  ARP 299, and fivefold system VV118, are correctly placed on Fig.1a, to avoid more cases of an incorrect usage in the literature. 

                                                                         Panoramic spectroscopy of MRK 171 was carried out on the base of observations with the 2.6-m telescope of Byurakan Astrophysical Observatory, using "VAGR" multipupil spectrograph in combination with CCD  Lick 2063x2058. For the both observations made with the different interference filters on the 13th and 16th of December 2004 dispersion of spectra in matrixes amounted 2.1/pix. There are two nebular emission lines of oxygen doublet [OIII]4959,5007 in the first spectral range, 490-520nm. In the second range, 640-680nm, among more intensive lines are Ha of hydrogen, and doublets [NII]6548,6583 and [SII] 6716,6731 of nitrogen and sulfur, respectively.  The both observational sets were incomplete, so that mainly relative intensities can be now used for analysis. 
        Resulting images, covering a round field of view of about 36 ang.s. in diameter with sample step size of 0.9 ang.s, contain almost the whole galaxy.
 
\section{Reducing and Results.}For reducing and visualization procedures of obtained data cubes the ADHOC software package, developed by J. Boulesteix, has been mainly used. Intensity maps have been constructed for separate emission lines. On Ha-image, Fig.1b, three well shaped condensations are seen within the component W. In the component E there are roughly 5 diffuse condensations, all with peak intensities fainter than those of condensations in W.  If  we look through the SDSS images of Mrk 171 for more morpfological details, the structure of the E component looks very similar, with some fainter condensations. At the same time some doubts arise on appearance of  the component W, as condensation W1 looks very bright in DR7, blending with W2 and W3, and conversely, too faint in later releases, till DR10.        
        
 A distribution of intensity ratio [NII]6583/Ha over the field was examined for a value close to 0.7, as a sign of AGN presence.The values of this ratio, averaged  for the separate  condensations, are given alongside with their designations in Figure 1b. They occupy the range 0.31\textless [NII]6583/Ha\textless0.47. Is noteworthy increase of the ratio along the structural chains, especially in the eastern component, from E1 to E4. In any case, the  structures differentiated in the two components of  the galaxy Mrk 171 are composed of HII-regions, i.e the regions exposed to processes only of starforming  activity.  \\
\textbf{\ Acknowledgements.}   Funding for SDSS-III has been provided by the Alfred P. Sloan Foundation, the Participating Institutions, the National Science Foundation, and the U.S. Department of Energy Office of Science. The SDSS-III web site is http://www.sdss3.org/.

\end{document}